\documentclass[aps,10pt,prl,superscriptaddress,showpacs,floatfix,twocolumn,longbibliography]{revtex4-1}

\usepackage{graphicx}
\usepackage{epstopdf}
\usepackage{amsmath}
\usepackage{amssymb}
\usepackage{hyperref}
\usepackage{bm}
\usepackage{subfigure}

\hypersetup{
    colorlinks=true,       
    linkcolor=cyan,          
    citecolor=magenta,        
    filecolor=magenta,      
    urlcolor=cyan,           
    runcolor=cyan
}
\usepackage[pdftex]{color}

\newcommand{\beq}{\begin{equation}}
\newcommand{\eeq}{\end{equation}}
\newcommand{\beqnn}{\begin{equation*}}
\newcommand{\eeqnn}{\end{equation*}}
\newcommand{\bea}{\begin{eqnarray}}
\newcommand{\eea}{\end{eqnarray}}
\newcommand{\beann}{\begin{eqnarray*}}
\newcommand{\eeann}{\end{eqnarray*}}
\newcommand{\bes} {\begin{subequations}}
\newcommand{\ees} {\end{subequations}}

\newcommand{\braket}[2]{\langle #1 | #2\rangle}
\newcommand{\ket}[1]{ | #1\rangle}

\newcommand{\ketbra}[2]{|#1\rangle\langle #2|}

\newcommand{\ignore}[1]{}
\newcommand{\GS}{0}

\begin{document}
\title{Temperature scaling law for quantum annealing optimizers}
\author{Tameem Albash}
\affiliation{Information Sciences Institute, University of Southern California, Marina del Rey, California 90292, USA}
\affiliation{Department of Physics and Astronomy and Center for Quantum Information Science \& Technology, University of Southern California, Los Angeles, California 90089, USA}

\author{Victor Martin-Mayor}
\affiliation{Departamento de F\'isica Te\'orica I, Universidad Complutense, 28040 Madrid, Spain}
\affiliation{Instituto de Biocomputaci\'on y F\'isica de Sistemas Complejos (BIFI), Zaragoza, Spain}

\author{Itay Hen}
\affiliation{Information Sciences Institute, University of Southern California, Marina del Rey, California 90292, USA}
\affiliation{Department of Physics and Astronomy and Center for Quantum Information Science \& Technology, University of Southern California, Los Angeles, California 90089, USA}
\email{itayhen@isi.edu}
\begin{abstract}
Physical implementations of quantum annealing unavoidably operate at finite temperatures. We {point to a fundamental limitation of} fixed finite temperature {quantum annealers that prevents them} from functioning as competitive scalable optimizers and show that to serve as optimizers annealer temperatures must be appropriately scaled down with problem size. We derive a temperature scaling law dictating that temperature must drop at the very least in a logarithmic manner but also possibly as a power law with problem size. We corroborate our results by experiment and simulations and discuss the implications of these to practical annealers.
\end{abstract}

\maketitle
%

{\bf Introduction}.--- {Quantum computing devices are becoming sufficiently large to undertake computational tasks that are infeasible using classical computing \cite{Bremner2017achievingsupremacy,PhysRevLett.118.040502,PhysRevLett.117.080501,PhysRevLett.112.130502,Broome794,Spring798,2016arXiv160800263B}.  
The theoretical underpinning for whether such tasks exist with physically realizable quantum annealers remains lacking, despite} the excitement brought on by recent technological breakthroughs that have made programmable quantum annealing (QA)~\cite{finnila_quantum_1994,Brooke1999,kadowaki_quantum_1998,farhi_quantum_2000,Santoro} optimizers consisting of thousands of quantum bits commercially available.
{Thus far, no examples of practical relevance have been found to indicate a superiority of QA optimization}, i.e., to find bit assignments that minimize the energy, or cost, of discrete combinatorial optimization problems, faster than possible classically~\cite{Young2008,Young:2010,hen:11,farhi:12,speedup,Hen:2015rt,q-sig,q-sig2}.
{Major ongoing efforts continue} to build larger, more densely connected QA devices, in the hope that the capability to embed larger optimization problems would eventually reveal the coveted quantum speedup~\cite{QEO,LL1,LL2,LL3,DW2000}.

{Understanding the robustness of QA optimization to errors that reduce the final ground state probability is critical.  In this work, we consider perhaps the most optimistic setting where the only source of error is due to nonzero temperature.} We analyze the theoretical scaling performance of \emph{ideal fixed-temperature} quantum annealers for optimization.  
We show that even in the case where annealers are assumed to \emph{thermalize instantly} (rather than only in the infinite runtime limit), the energies, or costs, of their output configurations would be computationally trivial to achieve (in a sense that we explain).  We further derive a scaling law for QA optimizers and provide corroboration of our analytical findings by experimental results obtained from the commercial D-Wave 2X  QA processor~\cite{Johnson:2010ys,Berkley:2010zr,Harris:2010kx,Bunyk:2014hb,King:2015}  as well as numerical simulations (our results equally apply to ideal thermal annealing devices).  We discuss the implications of our results for both {past benchmarking studies and for the engineering requirements of future QA devices}. 

{\bf Fixed-temperature quantum annealers}.---
In the adiabatic limit, closed-system quantum annealers are guaranteed to find a ground state of the target cost function, or final Hamiltonian $H$, they are to solve. The adiabatic theorem of quantum mechanics ensures that the overlap of the final state of the system with the ground state manifold of $H$, approaches unity as the duration of the process increases~\cite{Kato:50,Jansen:07}. 
For physical quantum annealers that operate at positive temperatures  ($T>0$), there is no equivalent guarantee of reaching the ground state with high probability. For long runtimes, an ideal finite-temperature quantum annealer is expected to sample the Boltzmann distribution of the final Hamiltonian at the annealer temperature~\cite{Venuti:2015kq}.

In what follows, we argue that even instantly-thermalizing quantum annealers \footnote{The existence of small minimum gaps prior to the end of the anneal suggests that it is extremely unlikely that the Gibbs state before crossing these gaps would have a larger overlap with the ground state manifold of the final Hamiltonian than after.  Therefore, measuring the system midway through a quantum annealing process will generically yield lower success probabilities than measurements taking place at the end of it~\cite{Altshuler13072010,farhi:12,knysh}.  We give two analytical examples in the Supplemental Information.} are severely limited as optimizers due to their finite temperature. 
For concreteness, we restrict to  
annealers for which i) the number of couplers scales linearly with the number of qubits $N$~\footnote{This is equivalent to having a bounded degree connectivity graph.}, ii) the coupling strengths are discretized and are bounded independently of problem size, and iii) the scaling of the free energy with problem size is not pathological, i.e., that our system is not tuned to a critical point. 
Other than the above standard assumptions, our treatment is general (we discuss the performance of quantum annealers when some of these conditions are lifted later on).  For clarity, we consider optimization problems written in terms of a Hamiltonian of the Ising-type
\beq \label{eqt:Ising}
H=\sum_{\langle i j \rangle} J_{ij} s_i s_j +\sum_i h_i s_i\,,
\eeq
where $\{s_i=\pm 1\}$ are binary Ising spin variables that are to be optimized over, $\{J_{ij},h_i\}$ are the coupling strengths between connected spins and external biases, respectively, and $\langle i j\rangle$ denotes the underlying connectivity graph of the model. The discussion that follows however is not restricted to any particular model.

Under the above assumptions, the ground state energies, denoted $E_{0}$, of any given problem class, scale linearly with increasing problem size (i.e., the energy is an extensive property as is generically expected from physical systems) while the classical minimal gap $\Delta=E_1-E_0$ remains fixed.
It follows then~\footnote{The analysis is based on the equivalence between the Canonical and the Microcanonical Ensembles of Statistical Mechanics. This equivalence is reviewed in many places, see e.g., Ref.~\cite{martin-mayor:07}.} that the thermal expectation values of the intensive energy
\beq
\langle e \rangle_\beta = \langle H \rangle_\beta / N \,,
\eeq
and specific heat
\beq
c_\beta = \partial \langle e \rangle_\beta / \partial \beta = -N \left[ \langle e^2 \rangle_{\beta} - \langle e \rangle^2_{\beta} \right]\,,
\eeq 
remain finite as $N \to \infty$ for any fixed inverse-temperature $\beta=1/T$.
The intensive energy is discretized in steps of $\Delta/N$, yet its statistical dispersion $\sigma_\beta(e)=\sqrt{-c_\beta/N}$ is much larger. Treating $e$ as a stochastic variable, for large enough values of $N$ it can be treated as a continuous variable as the ratio of discretization versus
dispersion is {$\sqrt{-\Delta^2/(c_\beta N)}$} decaying to zero for large $N$. From the Boltzmann distribution it follows that the probability density of $e$ goes as
$
p_\beta(e) ={Z_\beta}^{-1}\text{e}^{N (s(e) -\beta e)}\,,
$
where  $Z_\beta=\sum_{n} g_n \mathrm{e}^{-\beta E_n}$ is the partition function, $g_n$ is the degeneracy of the $n$-th level, i.e., the number of
microstates with $H(\{s_i\})=E_n$, satisfying $2^N =\sum_{n\geq 0} g_n$,  and $s(e)$ is the entropy density~\footnote{Equivalently, $N s(e)$ is the logarithm of the number of microstates with intensive energy $e$.}.
The linear combination $\Psi_\beta(e)=s(e) -\beta e$ 
plays the role of a large-deviations functional for $e$. The most probable value of $e$, which we denote by $e^\ast$, is given by the maximum of $\Psi_\beta$. Solving $\Psi_\beta'(e^*)=0$, we find~\footnote{A relation best known as the second law of thermodynamics {$\mathrm{d} e=T \mathrm{d} s$}.}
\beq
\beta = \left.\frac{\partial s}{\partial e}\right|_{e=e^*}\,.
\eeq
Close to $e^*$, $\Psi_\beta$ can be Taylor-expanded as
$
\Psi_\beta(e) \approx \Psi_\beta(e^*) - \frac{|\Psi_\beta''(e^*)|}{2}(e-e^*)^2
$, from which it follows that
\beq \label{eqt:Gaussian}
p_\beta(e)
\approx {\frac{\text{e}^{N \Psi_\beta(e^*)}}{Z_\beta} }\text{exp}\left[-\frac{N |\Psi_\beta''(e^*)|}{2}(e-e^*)^2 \right]\,.
\eeq
The probability density is thus approximately Gaussian in the vicinity of 
$e^*$, {although deviations from the Gaussian behavior are crucial} \footnote{{An energy probability density that is precisely Gaussian implies that the energy density is a linear function of the inverse temperature $\beta$ and hence the specific heat is a constant.  We elaborate on this point in the Supplemental Information.}}. Moreover, in the limit of large $N$, we find
\beq
\langle e\rangle_\beta = e^* \quad \text{and} \quad c_\beta=\frac{-1}{|\Psi_\beta''(e^*)|}\,.
\eeq
Therefore, the probability of finding by Boltzmann-sampling any energy $e < e^*$ (equivalently,
$E < e^* N$) is exponentially suppressed in $N$, scaling in fact as $\text{exp}[-N
 (\,\Psi_\beta(e^*)-\Psi_\beta(e)\, )]$.  
We thus arrive at the conclusion that even ideal fixed temperature quantum annealers that thermalize instantaneously to the Gibbs state of the classical Hamiltonian are \emph{exponentially unlikely} to find the ground state since $e^\ast > e_0 \equiv E_{0}/N$. 

\begin{figure}[ht] 
\includegraphics[width=0.98\columnwidth]{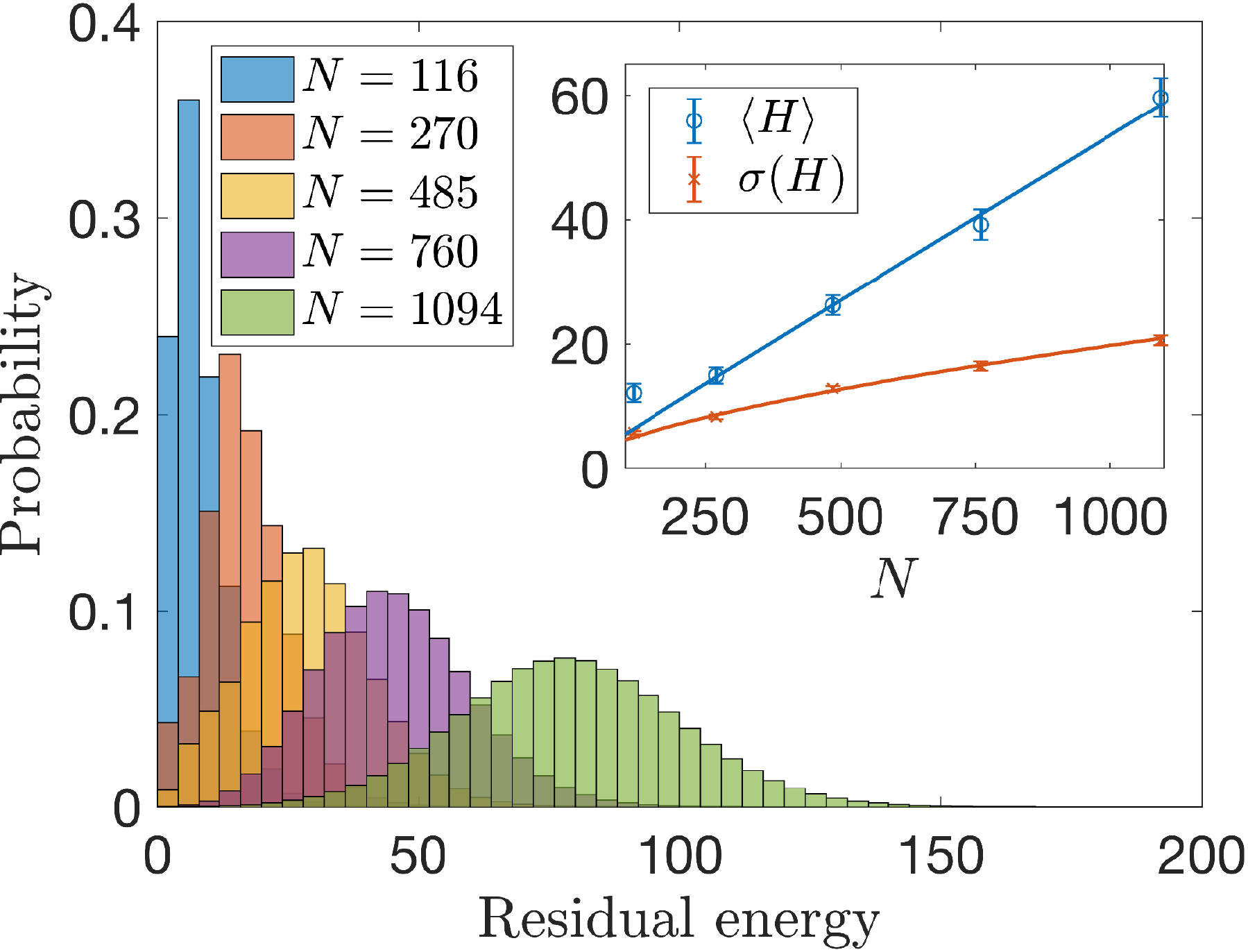}
   \caption{{\bf Distributions of residual energy, $E - E_{0}$, from DW2X runs.} As problem sizes grow, the distributions become more Gaussian-like. {\bf Inset:} 
   Gaussians' mean (blue) and standard deviation (red) as a function of problem size, averaged over 100 instances per size. The solid lines correspond to power-law fits  of the average mean with power $0.98 \pm  0.14$ and average standard deviation scaling with power $0.63 \pm 0.09$, taking into accounts all sizes but the smallest ($1.01 \pm 0.62$ and $0.57 \pm 0.37$ respectively if the two smallest sizes are omitted).}
   \label{fig:DW2XEnergyDistribution}
\end{figure}

We now corroborate the above derivation by runs on the commercial DW2X quantum annealer~\cite{Johnson:2010ys,Berkley:2010zr,Harris:2010kx,Bunyk:2014hb}. To do so, we first generate random instances of differently sized sub-graphs of the DW2X Chimera connectivity graph~\cite{Choi1,Choi2} and run them multiple times on the annealer, recording the obtained energies~\footnote{The reader is referred to the Supplemental Information for further details.}. 
Figure~\ref{fig:DW2XEnergyDistribution} depicts typical resultant residual energy ($E-E_0$) distributions. As is evident, increasing the problem size $N$ `pushes' the energy distribution farther away from $E_0$, as well as broadening the distribution and making it more gaussian-like.  
In the inset, we measure the departure of $\langle H \rangle_{\beta}$ from $E_0$ and the spread of the energies $\sigma_\beta(H)$ over 100 `planted-solution'~\cite{Hen:2015rt} instances per sub-graph size as a function of problem size $N$~\footnote{Details of these instances as well as similar results obtained for other problem classes are given in the Supplemental Information.}. For sufficiently large problem sizes, we find that the scaling of $\langle H -E_0\rangle_{\beta}$ is close to linear while $\sigma_\beta(H)$ scales slightly faster than $\sqrt{N}$.  While the slight deviations from our analytical predictions suggest that the DW2X configurations have not fully reached asymptotic behavior\footnote{The D-Wave processors are known to suffer from additional sources of error such as problem specification errors \cite{Martin-Mayor:2015dq,Zhu:2015pd} and freeze-out before the end of the anneal \cite{Amin:2015qf} that prevent thermalization to the programmed problem Hamiltonian.}, they exhibit a trend that closely matches our assumptions with the agreement getting better with growing problem sizes. 

Given the scaling of the mean and standard deviation, we conclude that fixed-temperature quantum annealers will generate energies $e$ with a fixed distance from $e_{0}$, or in terms of extensive energies, configurations obtained from fixed-temperature annealers will have energies concentrated around $E=(1-\epsilon) E_0$ for some $\epsilon>0$ {and $E_0 < 0$}.

One could now ask what the difficulty is for \emph{classical algorithms} to generate energy values in the above range.  
This question has been recently answered by the discovery of a polynomial time approximation scheme (PTAS) for spin-glasses defined on a Chimera graph
~\cite{PTAS_Saket} (and which can be easily generalized to any locally connected model), where reaching such energies can be done efficiently~\footnote{By no means however, is it meant that PTAS is able to thermally sample from a Boltzmann distribution of the input problem. In fact, it should be clear that PTAS does not.}. 
While the scaling of the PTAS with $\epsilon$ is not favorable, scaling as $c^{1/\epsilon}$ for some constant $c$, in practice there exist algorithms (e.g., parallel tempering that we discuss later on) that are known to scale more favorably than PTAS.
\\
{\bf Scaling law for quantum annealing temperatures}.---
In light of the above, it may seem that quantum annealers 
are doomed to fail as optimizers as problem sizes increase.
We now argue that success may be regained if the temperature of the QA device is appropriately scaled with problem size.  Specifically, we address the question of how the inverse-temperature $\beta$ should scale with $N$ such that there is a probability of at least $q$ of finding the ground state.  

An estimate for the required scaling can be given as follows.  From the above analysis, it should be clear that the probability of finding a ground state at inverse temperature $\beta$ will not decay exponentially with system size only if the ground state falls within the variation of the mean energy, specifically if
\beq
\sigma_{\beta}(H)= N \sigma_\beta(e)=\sqrt{-N c_\beta}\,,
\eeq
is comparable to
\beq
\langle H \rangle_\beta -E_0= - N \int_\beta^\infty\mathrm{d}\,\beta\ c_\beta\,.
\eeq
The third law of thermodynamics dictates that the specific heat $c_T\equiv \mathrm{d}\langle e\rangle/\mathrm{d} T$
goes to zero when $T\to 0$. Assuming a scaling of the form
$
c_T \sim T^\alpha
$, or equivalently, $-c_\beta\sim \beta^{-\alpha-2}$, gives
\beq
\sigma_{\beta}(H)\sim \sqrt{\frac{N}{\beta^{\alpha+2}}}\quad \text{and} \quad \langle
H\rangle_\beta -E_0= \frac{N}{\beta^{\alpha+1}}\,.
\eeq
For a power-law specific heat, it thus follows that the sought scaling is
$\beta\sim N^{1/\alpha}$.
If on the other hand $c_\beta$ vanishes exponentially in $\beta$, the inverse-temperature scaling will be milder, of the form $\beta\sim \log{N}$.  

To illustrate the above, we next present an analysis of simulations of randomly generated instances on Chimera lattices (we study several problem classes and architectures, see the Supplemental Information).  To study the energy distribution generated by a thermal sampler on these instances, we use parallel tempering (PT)~\cite{Geyer:91,Hukushima:1996}, a Monte Carlo method whereby multiple copies of the system at different temperatures are simulated~\footnote{Details of our PT implementation can be found in the Supplemental Information.}. 
In Fig.~\ref{fig:PTEnergyDistribution}, we show an example  of how the energy distribution of a planted-solution instance changes with $\beta$.  The qualitative behavior is similar to what we observe with increasing problem size, whereby decreasing $\beta$ (increasing the temperature) pushes the energy distribution to larger energies and makes it more gaussian-like.
\begin{figure}[ht] 
\includegraphics[width=0.98\columnwidth]{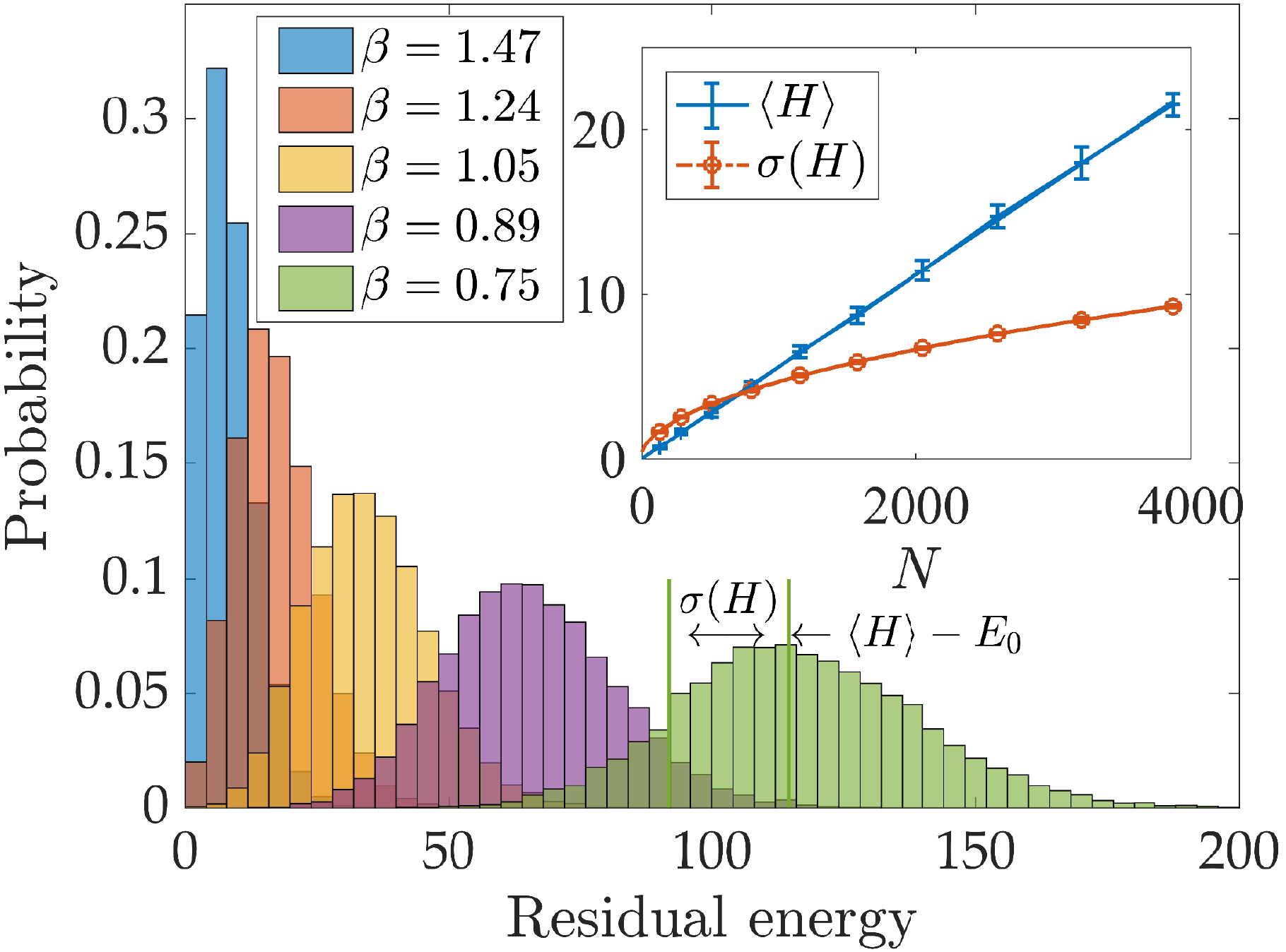}
   \caption{{\bf Distributions of residual energy, $E - E_{0}$, from PT simulations.} For a planted-solution instance defined on an $L=12$ Chimera graph, the distributions become more Gaussian-like as $\beta$ decreases.  For the case of $\beta =0.75$, the mean residual energy and standard deviation are indicated. {\bf Inset:} Scaling with problem size of the median mean energy and median standard deviation of the energy for $\beta = 1.47$  over 100 instances.}\label{fig:PTEnergyDistribution}
\end{figure}

\begin{figure}[htp] 
   \centering
   {\includegraphics[width=0.8\columnwidth]{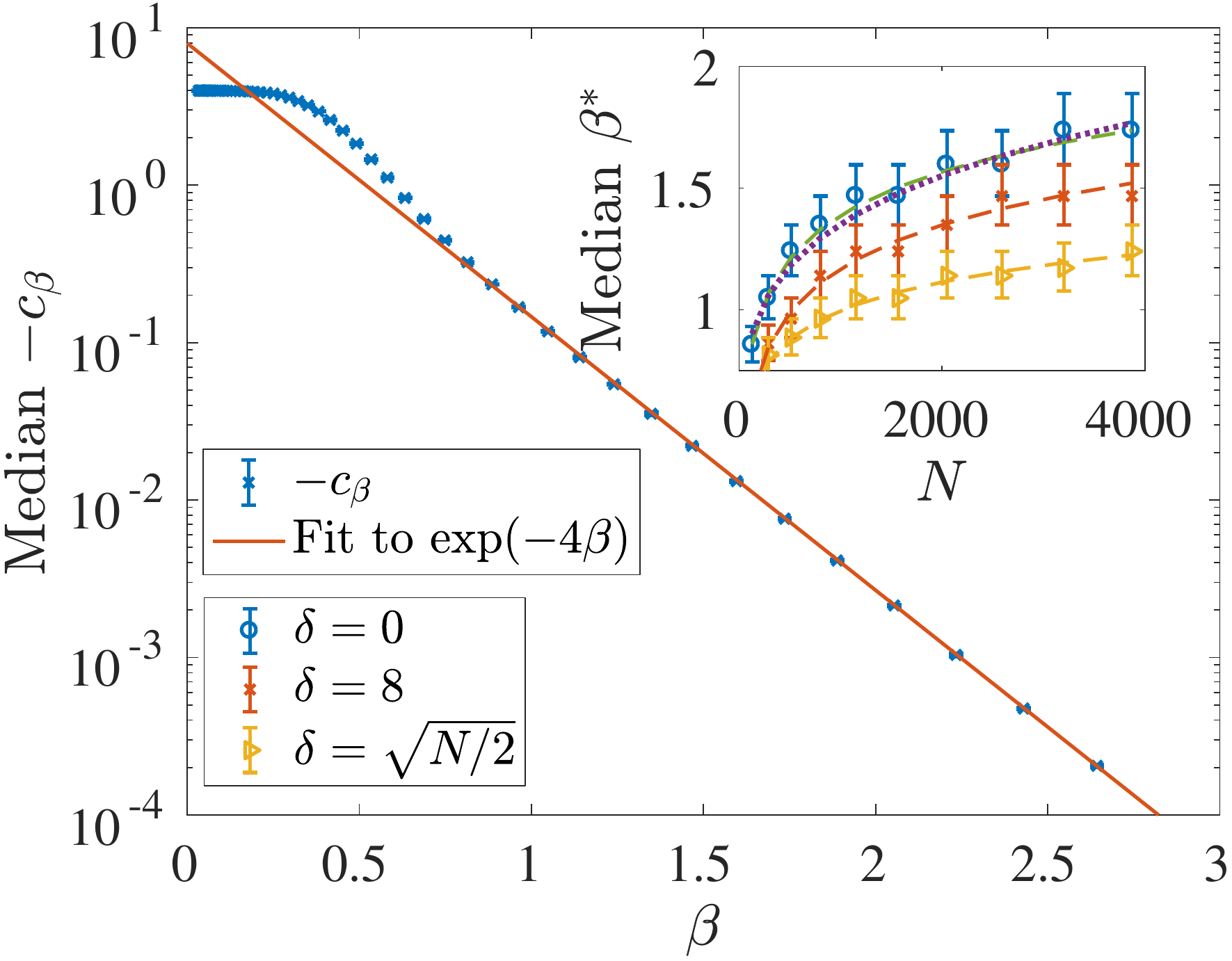} }
   \caption{{\bf Typical specific heat with inverse-temperature}. Behavior of the median specific heat (over 100 instances) for planted-solution instances with inverse-temperature $\beta$ for $N =3872$. The behavior transitions from a polynomial scaling with $\beta$ to an exponential scaling. {\bf Inset:} Typical minimum inverse-temperature required for instances of size $N$ such that the probability of the target energy $E_{\mathrm{T}} = E_{0} + \delta(N)$ is at least $q = 10^{-1}$. Also shown are fits to $\log N$ for all three cases and a power-law fit to $c N^{\alpha}$ that finds $\alpha=0.19\pm0.05$ for the $\delta=0$ case, which is almost indistinguishable from the logarithmic fit.}
\label{fig:BetaTarget}
\end{figure}

The behavior of the specific heat $c_{\beta}$ as the inverse-temperature $\beta$ becomes large is shown in Fig.~\ref{fig:BetaTarget}.  
At large sizes, the scaling becomes $c_\beta \propto \exp(-\Delta \beta )$ as expected (here, $\Delta=4$ is the gap).  Based on our predictions above, this should mean that if for a fixed $q$, the minimum $\beta^\ast$ such that $p_{\beta^\ast}(E_0) \geq q$ falls in this exponential regime, then we should observe a scaling $\beta^\ast \propto \log N$.  Indeed, the inset of Fig.~\ref{fig:BetaTarget}, which shows simulation results of $\beta^\ast$ versus $N$,  exhibits the expected $\log N$ behavior~\footnote{Similar scaling behavior for other classes of Hamiltonians, specifically 3-regular 3-XORSAT instances and random $\pm 1$ instances, is also observed, and we give the results for these instances in the Supplemental InformationSupplemental Information.}.

While for problem classes with a fixed minimum gap $\Delta$, one may naively expect $c_\beta$ to vanish exponentially in general, implying that a logarithmic scaling of $\beta$ will generally be sufficient as our simulations indeed indicate, it is important to note that two-dimensional spin glasses are known to exhibit a crossover between an exponential behavior to a power law~\cite{jorg:06b,thomas:11,parisen:11,fernandez:16}.  This crossover is characterized by a constant $\theta\approx 1/2$, whereby the discreteness of the gap $\Delta$ is evident only for sizes $N^{\theta/2} \ll \beta$. Beyond $N^{\theta/2} \sim \beta$, the $2d$ system behaves as if the coupling distribution is continuous~\cite{thomas:11,parisen:11} at which point the system can be treated as if with continuous couplings, for which the specific heat $c_T$ scales as $T^{\alpha}$ with $\alpha_c=2\nu$~\cite{jorg:06b}, where $\nu=3.53(7)$~\cite{fernandez:16}.  Therefore, for an ideal quantum annealer operating beyond the crossover, a scaling of $ \beta \sim N^{1/(2\nu) \approx 0.14}$ is required. We may thus expect the same crossover to appear for instances defined on the Chimera lattice, which is $2d$-like. Interestingly, for the temperature scaling shown in the inset of Fig.~\ref{fig:BetaTarget}, a power-law fit  $\beta \sim N^{\alpha}$ with $\alpha= 0.19 \pm 0.05$ is almost indistinguishable from the logarithmic one, with a power that is consistent with the $2d$ prediction.

{\bf Suboptimal metrics for optimization problems}.---
For many classically intractable optimization problems, when formulated as Ising models, it is crucial that solvers find a true minimizing bit assignment rather than low lying excited states.  This is especially true for NP-complete/hard problems~\cite{2013arXiv1302.5843L} where sub-optimal costs generally correspond to violated constraints that \emph{must} be satisfied (otherwise the resultant configuration is nonsensical despite its low energy). Nonetheless, it is plausible to assume the existence of problems for which slightly sub-optimal configurations would still be of value~\cite{Vinci:2016}.  We thus also study the necessary temperature scaling for cases where the target energies obey $E_{\mathrm{T}}\leq E_0+\delta(N)$ with $\delta(N)$ scaling sub-linearly with problem size. In the inset of Fig.~\ref{fig:BetaTarget}, we plot the required scaling of $\beta$ for  $\delta(N) = \mathrm{const}$ and $\delta(N) \propto \sqrt{N}$. In both cases we find that a logarithmic scaling is still essential, albeit with smaller prefactors.

{\bf Conclusions and discussion}.---
We have shown that fixed temperature quantum annealers can only sample `easily reachable' energies in the large problem size limit, {thereby posing fundamental limitation on their performance}.  
We derived a temperature scaling law to ensure that quantum annealing optimizers find nontrivial energy values with sub-exponential probabilities. The scaling of the specific heat with temperature controls this scaling: if $\beta$ lies in the regime where the specific heat scales exponentially with $\beta$, then the inverse-temperature of the annealer must scale as $\log N$.  However, further considerations are needed because of a possible crossover behavior in the specific heat with temperature and problem size. 
For Chimera graphs, because of their essentially two-dimensional structure, this may lead to a crossover to power law scaling.
 Little is known about this crossover in three dimensions or for different architectures, so this concern may not be mitigated by a more complex connectivity graph.     

Our results shed important light on benchmarking studies that have found no quantum speedups~\cite{q108,speedup,Hen:2015rt,Martin-Mayor:2015dq,Zhu:2015pd}, identifying temperature as a relevant culprit for their unfavorable performance. Our analysis is particularly relevant for both the utility as well as the design of future QA devices that have been argued to sample from thermal or close-to-thermal distributions~\cite{PhysRevA.92.052323}, calling their role as optimization devices into question. 

One approach to scaling down the temperature with problem size is the (theoretically) equivalent scaling up of the overall energy scale of the Hamiltonian. However, the rescaling of the total Hamiltonian is also known to be challenging and may not represent a convenient approach for a scalable architecture.  An alternative approach is to develop quantum error correction techniques to effectively increase the energy scale of the Hamiltonian by coupling multiple qubits to form a single logical qubit~\cite{PAL:13,PAL:14,Vinci:2015jt,MNAL:15,NQAC:2016,MNVAL:17} in conjunction with classical post-processing \cite{1367-2630-19-2-023024,2016arXiv160905875C,Karimi2017,2017arXiv170607826K} or to effectively decouple the system from the environment \cite{jordan2006error,Bookatz:2014uq,Jiang:2015kx,Marvian:2017}.

Our results reiterate the need for fault-tolerant error correction for scalable quantum annealing, however they do not preclude the utility of quantum annealing optimizers for large finite size problems, where engineering challenges may be overcome to allow the device to operate effectively at a sufficiently low temperature such that problems of interest of a finite size may be solved even in the absence of fault-tolerance.  Our results only indicate that this `window of opportunity' cannot be expected to continue as devices are scaled without further improvements in the device temperature or energy scale.

While our arguments above indicate that fixed-temperature quantum annealers may not be scalable as optimizers, 
the current study does not pertain to the usage of quantum annealers as \emph{samplers} ~\cite{PhysRevA.92.052323,Adachi:2015qe,Benedetti:2016}, where the objective is to sample from the Boltzmann distribution. The latter objective is known to be very difficult task (it is \#P-hard~\cite{Papadimitriou:book,Valiant:1979,Dahllof:2005}) and little is known about when or if quantum annealers can provide an advantage in this regard~\cite{2016arXiv161207979C}.

\begin{acknowledgments}
{\bf Acknowledgements}.---
TA and IH thank Daniel Lidar for useful comments on the manuscript. The computing resources were provided by the USC Center for High Performance Computing and Communications.  TA was supported under ARO MURI Grant No. W911NF-11-1-0268, ARO MURI Grant No. W911NF-15-1-0582, and NSF Grant No. INSPIRE- 1551064. V. M.-M. was partially supported by MINECO (Spain) through Grant No. FIS2015-65078- C2-1-P (this contract partially funded by FEDER). 
\end{acknowledgments}
%

%
\newpage
\onecolumngrid
\begin{center}
\textbf{\large{Supplemental Information for}}\\
\textbf{\large{``Temperature scaling law for quantum annealing optimizers"}}\\
\end{center}

\twocolumngrid
{\section{Deviations to the Gaussian probability density}
In the main text, we indicated that deviations from a Gaussian distribution for the marginal of the classical
Boltzmann probability (at inverse temperature $\beta$) for the energy density $p_{\beta}$ is crucial.  To see
why this is the case, let us consider what happens when the
probability density is \emph{exactly} Gaussian:
\begin{equation}\label{eq:Gauss-ansatz}
p_\beta(e)=\sqrt{\frac{N}{2\pi |c_\beta|}}\, \mathrm{e}^{-N\frac{[e-\langle e\rangle_\beta]^2}{2 |c_\beta|} }  \ .
\end{equation}
The probability density at any other inverse temperature $\beta +\delta \beta$ can be obtained as \cite{FALCIONI1982331,PhysRevLett.63.1195}
\begin{equation}\label{eq:reweighting}
p_{\beta+\delta\beta}(e)=\frac{1}{\cal Z} p_\beta(e)\, \mathrm{e}^{-N \delta\beta e}\,,\quad {\cal Z}=\int_{-\infty}^{\infty}\mathrm{d}e\  p_\beta(e)\, \mathrm{e}^{-N \delta\beta e}\,.
\end{equation}
Eq.~(\ref{eq:reweighting}) is fully general, and 
 we are not assuming $\delta\beta$ to be small. Now,
let us plug the Gaussian probability (Eq.~\eqref{eq:Gauss-ansatz}) into
Eq.~\eqref{eq:reweighting}. We find 
\begin{equation}\label{eq:bububu}
p_{\beta+\delta\beta}(e)=\frac{1}{\cal Z} \mathrm{exp}\big[-\frac{N}{2 |c_\beta|} [e -\langle e \rangle_\beta - c_\beta \delta\beta]^2\big] \ , 
\end{equation}
with ${\cal Z}= \sqrt{\frac{N}{2\pi |c_\beta|}} $.
Comparing Eqs.~\eqref{eq:Gauss-ansatz} and \eqref{eq:bububu}, we are led to the conclusion that the Gaussian probability implies that the energy density is a linear function of $\beta$ and that the specific heat is constant:
\begin{equation}\label{eq:Gauss-disaster}
\mathrm{Gaussian\ hypothesis:}\quad \langle e\rangle_\beta=\langle e\rangle_{\beta=0}+\beta c_\beta\,,\quad \frac{\mathrm{d} c_\beta}{\mathrm{d}\beta}=0\,.
\end{equation} 
}

{ Of course, Eq.~(\ref{eq:Gauss-disaster}) is grossly in
  error, because in the limit of large $\beta$ (i.e. zero temperature)
  $\langle e\rangle_\beta$ should reach the Ground State
  energy-density [rather than diverge as wrongy implied by
  Eq.~(\ref{eq:Gauss-disaster})]. In fact, the specific heat is not
  constant. A straightforward application of the
  fluctuation-dissipation theorem tells us that
\begin{equation}\label{eq:well_well_well}
\frac{\mathrm{d} c_\beta}{\mathrm{d}\beta} = N^2 \big[\langle e^3\rangle - 3 \big\langle e^2\rangle \langle e \rangle + 2 \langle e \rangle^3] = N^2 \big\langle \,[e -\langle e\rangle]^3\,\big\rangle\,.
\end{equation}
We can introduce $\eta$, the fluctuating part of the energy (regarded
as a stochastic variable):
\begin{equation}
e=\langle e \rangle_\beta + \sqrt{\frac{-c_\beta}{N}}\eta\,. 
\end{equation}
In combination with the fluctuation-dissipation theorem, $c_\beta=-N\langle [e -\langle e\rangle]^2\rangle$, we have
$$\langle\eta\rangle=0\,,\quad \langle \eta^2\rangle=1\,.$$
Furthermore, Eq.~(\ref{eq:well_well_well}) implies 
\begin{equation} \label{eq:eta3}
\langle \eta^3\rangle=\frac{1}{\sqrt{N}}\frac{1}{[-c_\beta]^{3/2}} \frac{\mathrm{d} c_\beta}{\mathrm{d}\beta}\,.
\end{equation}
However, if $\eta$ be a normal variable $N(0,1)$ as demanded by
Eq.~(\ref{eq:Gauss-ansatz}), we would have $\langle
\eta^3\rangle=0$ and not what we have in Eq.~\eqref{eq:eta3}. Hence, convergence to the main traits of the Gauss distribution
law (symmetry under $\eta \leftrightarrow -\eta$, for instance) happens at a rate proportional to $1/\sqrt{N}$.
}

\section{The DW2X experimental quantum annealing optimizer} \label{sec:DW2X}

\subsection{Description of the processor}

\begin{figure*}[htbp] 
   \centering
   \includegraphics[angle=270,width=1.85\columnwidth]{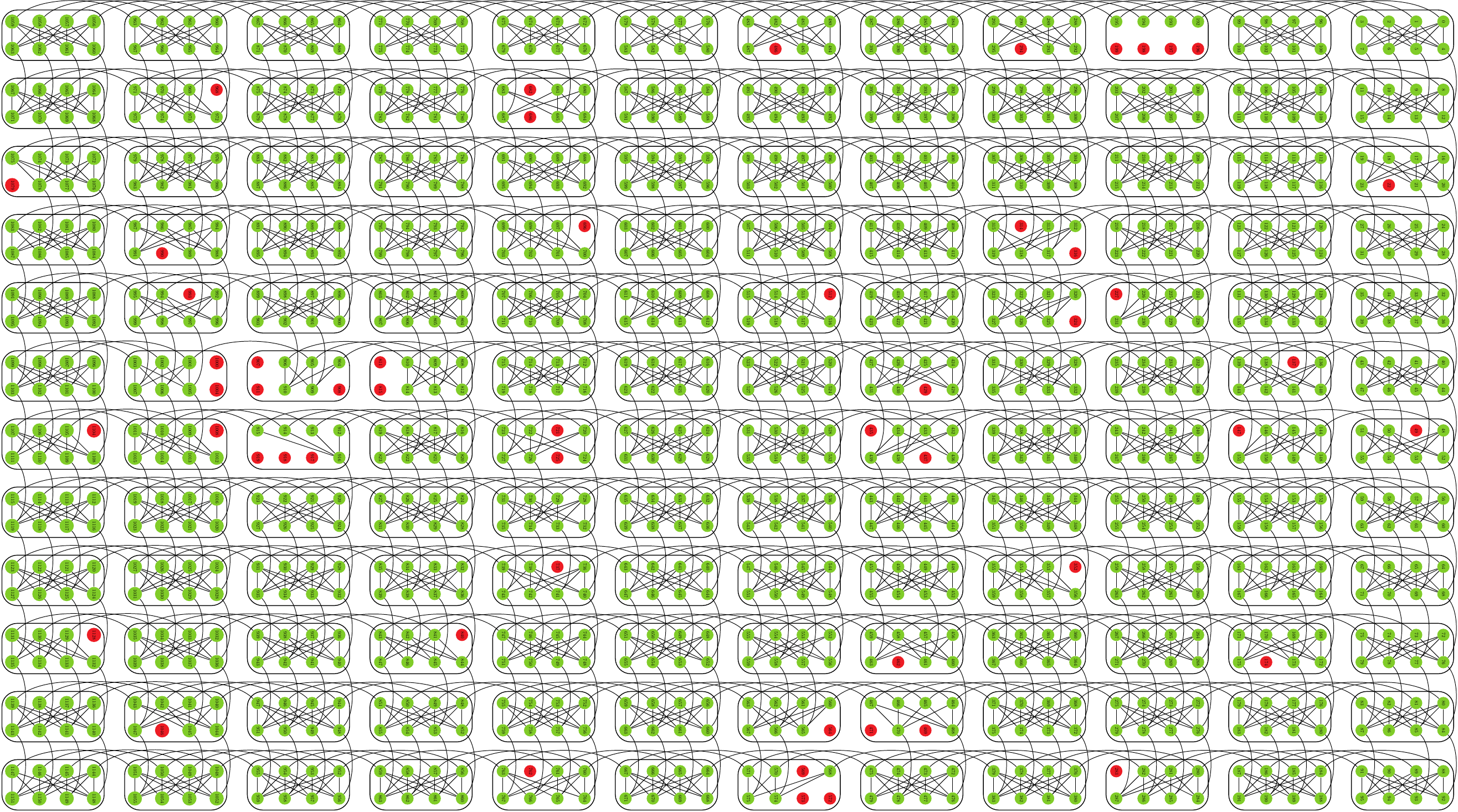} 
   \caption{{\bf A visualization of the DW2X graph.}  Operational qubits are shown in green, and inoperational ones are shown in red.  Programmable couplers are shown as black lines connecting the qubits.}
   \label{fig:DW2Xgraph}
\end{figure*}

The experimental results shown in the main text were taken on a 3rd generation D-Wave processor, the DW2X `Washington' processor, installed at the Information Sciences Institute - University of Southern California (ISI).  The processor connectivity is given by a $12 \times 12$ grid of unit cells, where each unit cell is composed of 8 qubits with a $K_{4,4}$ bipartite connectivity, forming the `Chimera' graph~\cite{Choi1,Choi2} with a total of 1152 qubits.  Due to miscalibration, there are only 1098 operational qubits on the ISI machine.  This is illustrated in Fig.~\ref{fig:DW2Xgraph}.  

\begin{figure}[htbp] 
   \centering
   \includegraphics[width=0.95\columnwidth]{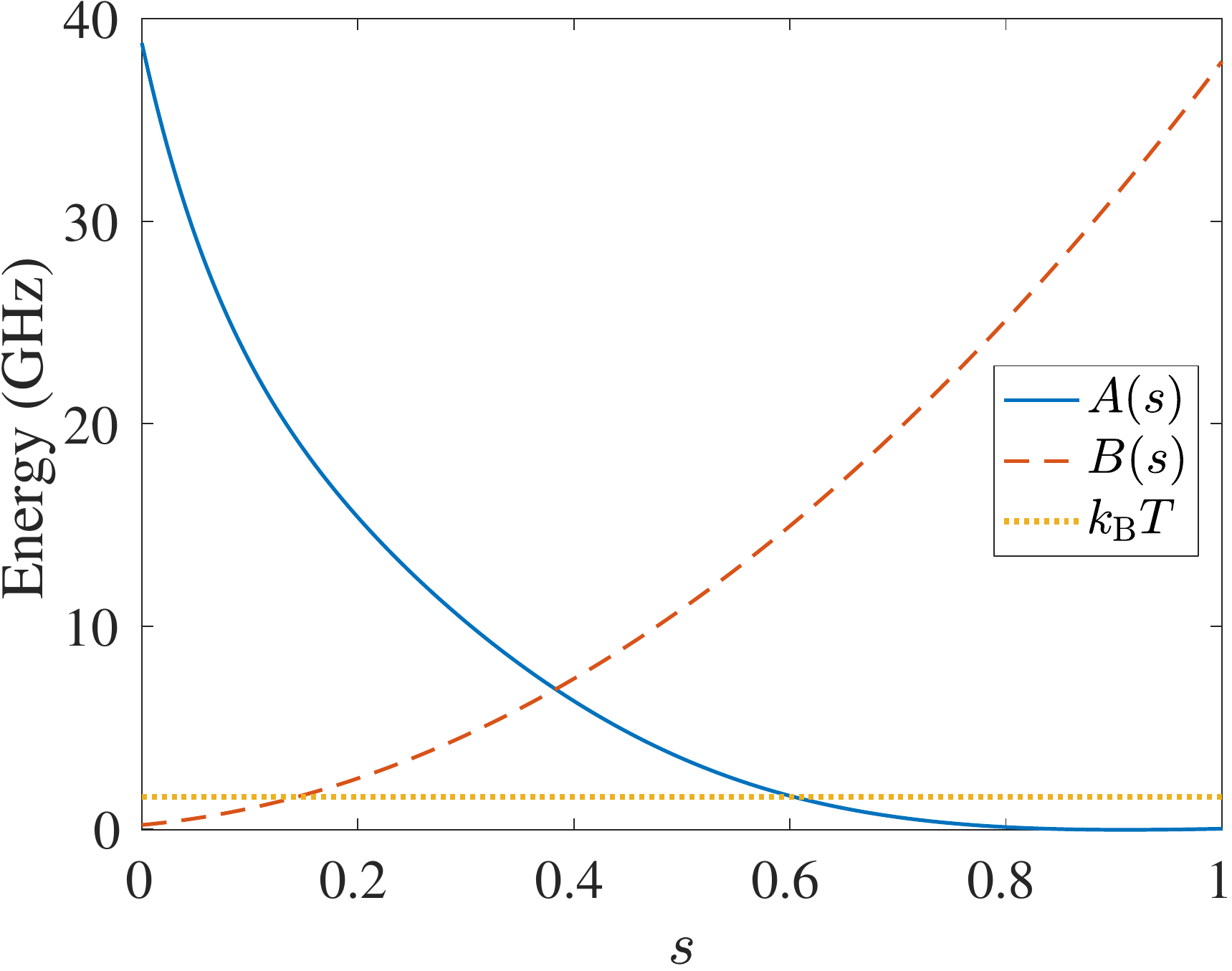} 
   \caption{{\bf The DW2X annealing schedules.}  Energy units are for $\hbar = 1$, and the operating temperature of $12m$K is shown as well.}
   \label{fig:DW2Xschedule}
\end{figure}

The device implements the quantum annealing protocol given by the time-dependent Hamiltonian:
\beq
H_{\mathrm{QA}}(s) = A(s) H_{\mathrm{D}} + B(s) H 
\eeq
where $H_{\mathrm{D}} = - \sum_i \sigma_i^x$ is the standard transverse field driver Hamiltonian, $H$ is the Ising Hamiltonian [Eq.~(1) of the main text], and $A(s), B(s)$ are the annealing schedules satisfying $A(0) \gg B(0)$, $A(1) \ll B(1)$, and $s \equiv t/t_f \in [0,1]$ is the dimensional time annealing parameter.  The predicted functional form for these schedules is shown in Fig.~\ref{fig:DW2Xschedule}. 

\subsection{Details of the experiment and additional results}

\begin{figure}[ht] 
\includegraphics[width=0.98\columnwidth]{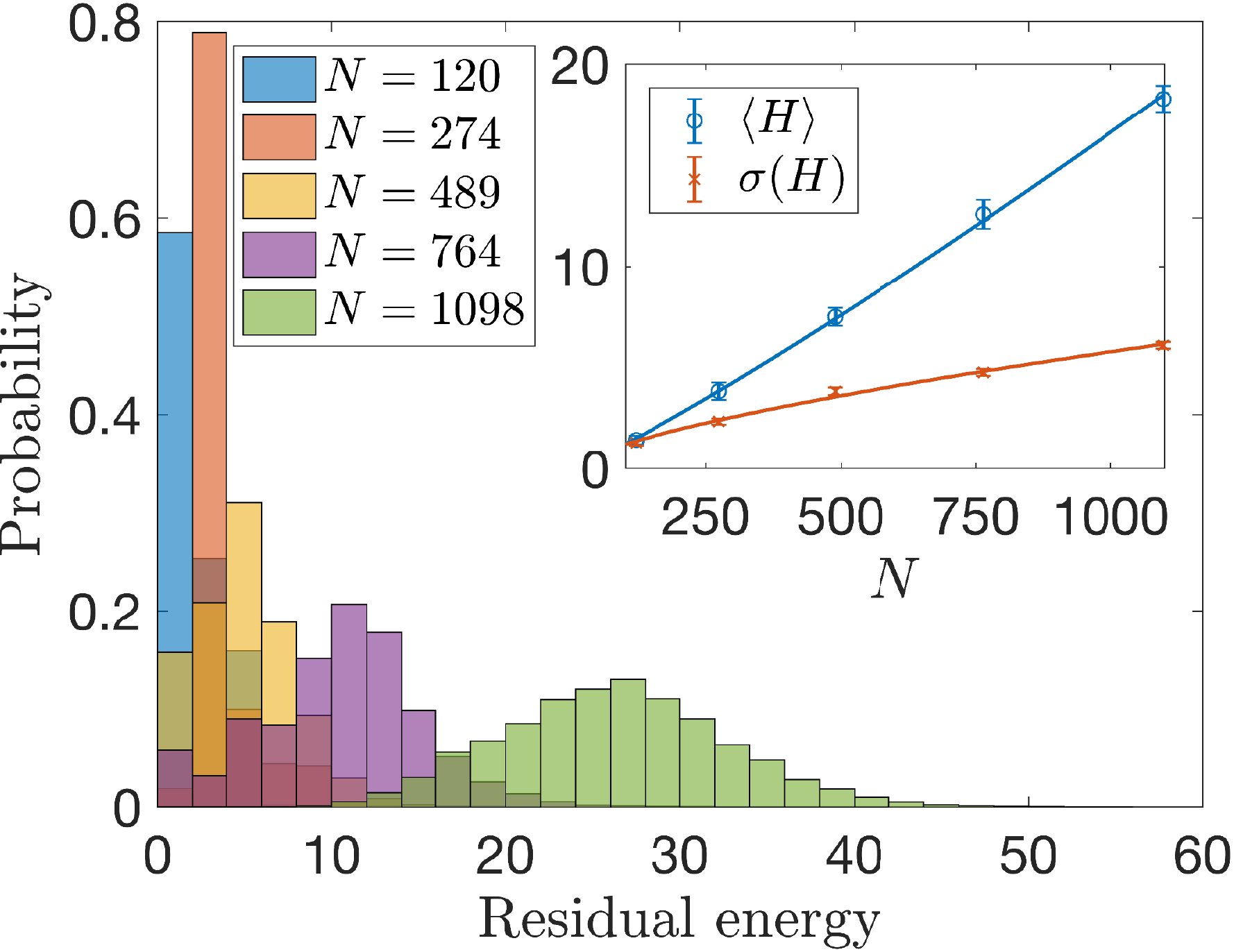}
   \caption{{\bf Distributions of residual energy, $E - E_{0}$, from DW2X simulations on random $\pm 1$ instances.} As problem sizes grow, the distributions become more Gaussian-like. {\bf Inset:} 
   Gaussians' mean (blue) and standard deviation (red) as a function of problem size, averaged over 100 instances per size. The solid lines correspond to best fits to the form $\ln(y) = a + b \ln(x)$, with $a = -5.00 \pm 0.53, b=1.13 \pm  0.08$ and $a = -2.92 \pm 0.37 , b = 0.68 \pm 0.06 $ respectively, taking into accounts all sizes but the smallest.
   }
   \label{fig:DW2XEnergyDistributionBimodal}
\end{figure}

The randomly generated instances tested on the D-Wave processor were run with 20 random gauges~\cite{q-sig} with 5000 reads per gauge/cycle for a total of $100,000$ anneals per instance. The annealing time chosen for the runs was the default $20 \mu$-sec. 
We further corroborated the analytical derivations discussed in the main text using experiments on the commercial DW2X processor on randomly generated bi-modal $J_{ij}=\pm 1$ instances. As with the planted-solution instances, we first generate random instances of differently sized sub-graphs of the DW2X Chimera connectivity graph~\cite{Choi1,Choi2} and run them multiple times on the annealer, recording the obtained energies. 
Figure~\ref{fig:DW2XEnergyDistributionBimodal} depicts the resultant residual energy ($E-E_0$) distributions of a typical instance. As is evident, increasing the problem size $N$ `pushes' the energy distribution farther and farther away from the ground state value, as well as broadening the distribution and making it more gaussian-like.  
In the inset we measure the departure of $\langle H \rangle_{\beta}$ from $E_0$ and the spread of the energies $\sigma_\beta(H)$ over 100 random bi-modal instances per sub-graph size as a function of problem size $N$. For sufficiently large problem sizes, we find that the scaling of $\langle H -E_0\rangle_{\beta}$ is almost linear while $\sigma_\beta(H)$ scales slightly faster than $\sqrt{N}$.  The results are slightly worse than the analytical prediction but conform to the general trend.

\section{Simulation Methods}

\subsection{Instance generation} \label{sec:Instances}
For the generation of instances in this work we have chosen one problem class to be that of the `planted solution' type---an idea borrowed from constraint satisfaction (SAT) problems. In this problem class, the planted solution represents a ground-state configuration of the Hamiltonian that minimizes the energy and is known in advance. The Hamiltonian of a planted-solution spin glass is a sum of terms, each of which consists of a small number of connected spins, namely, $H=\sum_j H_j$\cite{Hen:2015rt}. Each term $H_j$ is chosen such that one of its ground-states is the planted solution. It follows then that the planted solution is also a ground-state of the total Hamiltonian, and its energy is the ground-state energy of the Hamiltonian. 
Knowing the ground-state energy in advance circumvents the need to verify the ground-state energy using exact (provable) solvers, which rapidly become too expensive computationally as the number of variables grows. The interested reader will find a more detailed discussion of planted Ising problems in Refs.\cite{Hen:2015rt,King:2015zr}. 

For the random $\pm 1$ instances on Chimera, we randomly (with equal probability) assign a value $\pm 1$ to all the edges of the Chimera graph.  While the ground state energy for these instances is not known with 100\% certainty, we ran the Hamze-Freitas-Selby algorithm (HFS)~\cite{hamze:04,Selby:2014tx} for a sufficiently long time such that we were confident of having found the ground state for these instances.
   
For the 3-regular 3-XORSAT instances, for each spin, we randomly pick three other spins to which to couple.  All couplings are picked to be antiferromagnetic with strength $1$.  Because all terms in the Hamiltonian are of the form $+\sigma_i^z \sigma_j^z \sigma_k^z$, the ground state is simply that all-spins-down state.
\subsection{Parallel tempering} \label{sec:PT}
\begin{figure}[htbp] 
   \centering
   \includegraphics[width=0.95\columnwidth]{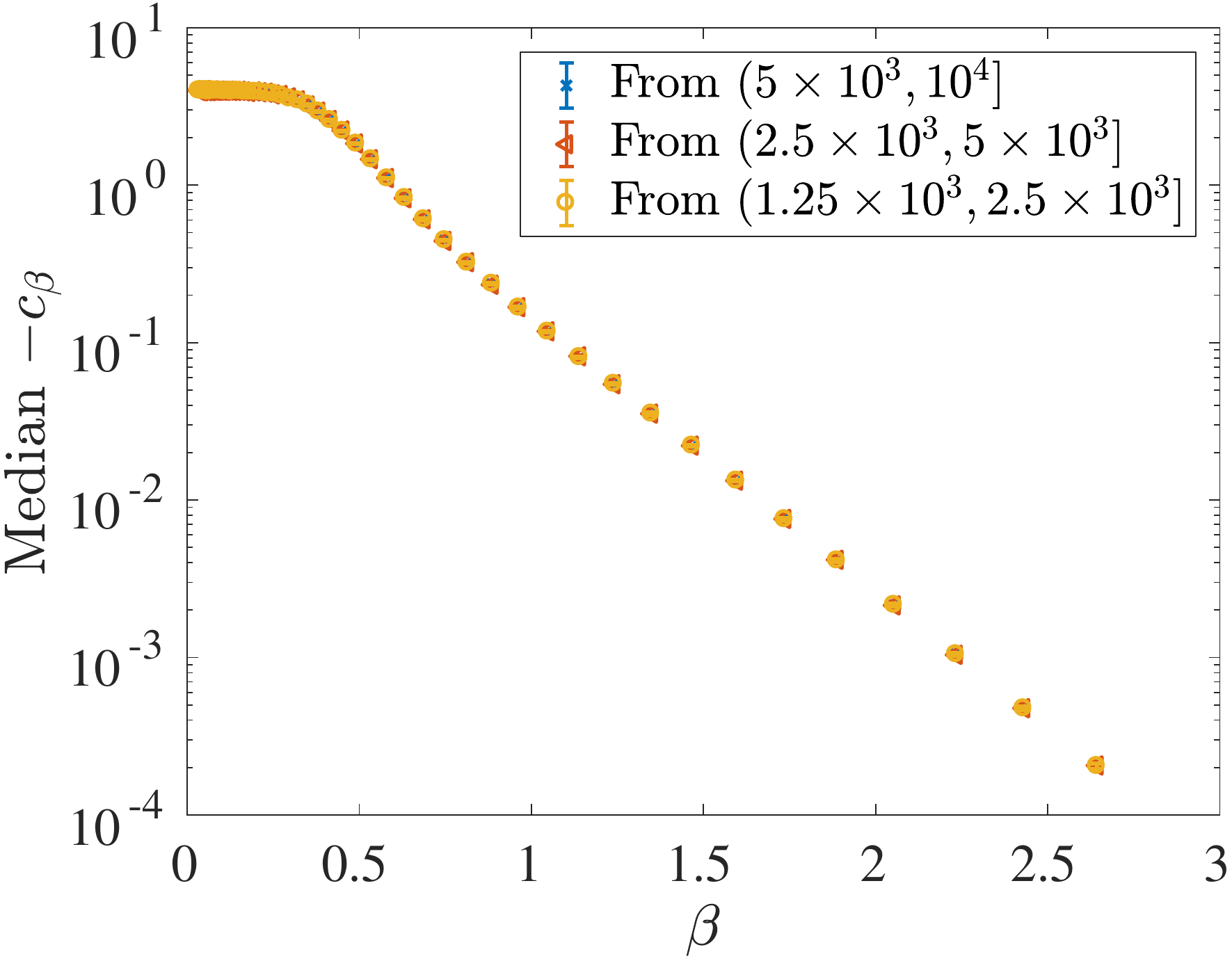} 
   \caption{{\bf The behavior of the median specific heat for the planted-solution instances at $L=22$ using different blocks of the samples}.  Of the total of $10^4$ samples, different partitions (as indicated by the legend) are used to calculate the specific heat.}
   \label{fig:PlantedTest}
\end{figure}

%
For the planted-solution instances, we first `warmed-up' our parallel tempering simulation with $5\times 10^5$ (for the smaller sizes) to $2\times 10^6$ (for the larger sizes) swaps with 10 Monte Carlo sweeps per swap.  The temperature distribution is picked as follows:
\beq
\beta_{i} = \left( \frac{\beta_{63}}{\beta_0} \right)^{i/63} \beta_0 \ , \quad i = 0, 1, \dots, 63
\eeq
with $\beta_0 = 20$ and $\beta_{63} = 0.1$.  After the warm-up, we sample the energy after every 50 swaps in order to minimize correlation between the energies.  We use a total of $10^4$ sample points, from which we extract the energies at different quantiles. In order to ensure that we have reached a thermal or near-thermal distribution, we performed the following check.  The $10^4$ sample points are divided into three blocks: (a) $5 \times 10^3$ samples from the last half of the samples; (b) $2.5 \times 10^3$ samples from the second quarter of the samples; (c) $1.25\times 10^3$ samples from the second eighth of the samples.  We then calculated the specific heat using the samples from each block separately; if the system has sufficiently thermalized and the samples are sufficiently uncorrelated, we expect to observe no change in the specific heat for the three sets of samples within the error bars.  We show the results of this test in Fig.~\ref{fig:PlantedTest}, where we indeed observe no significant difference.

\section{Results for Planted-solution Instances with a Target Energy}
\begin{figure}[htbp] 
   \centering
   \includegraphics[width=0.95\columnwidth]{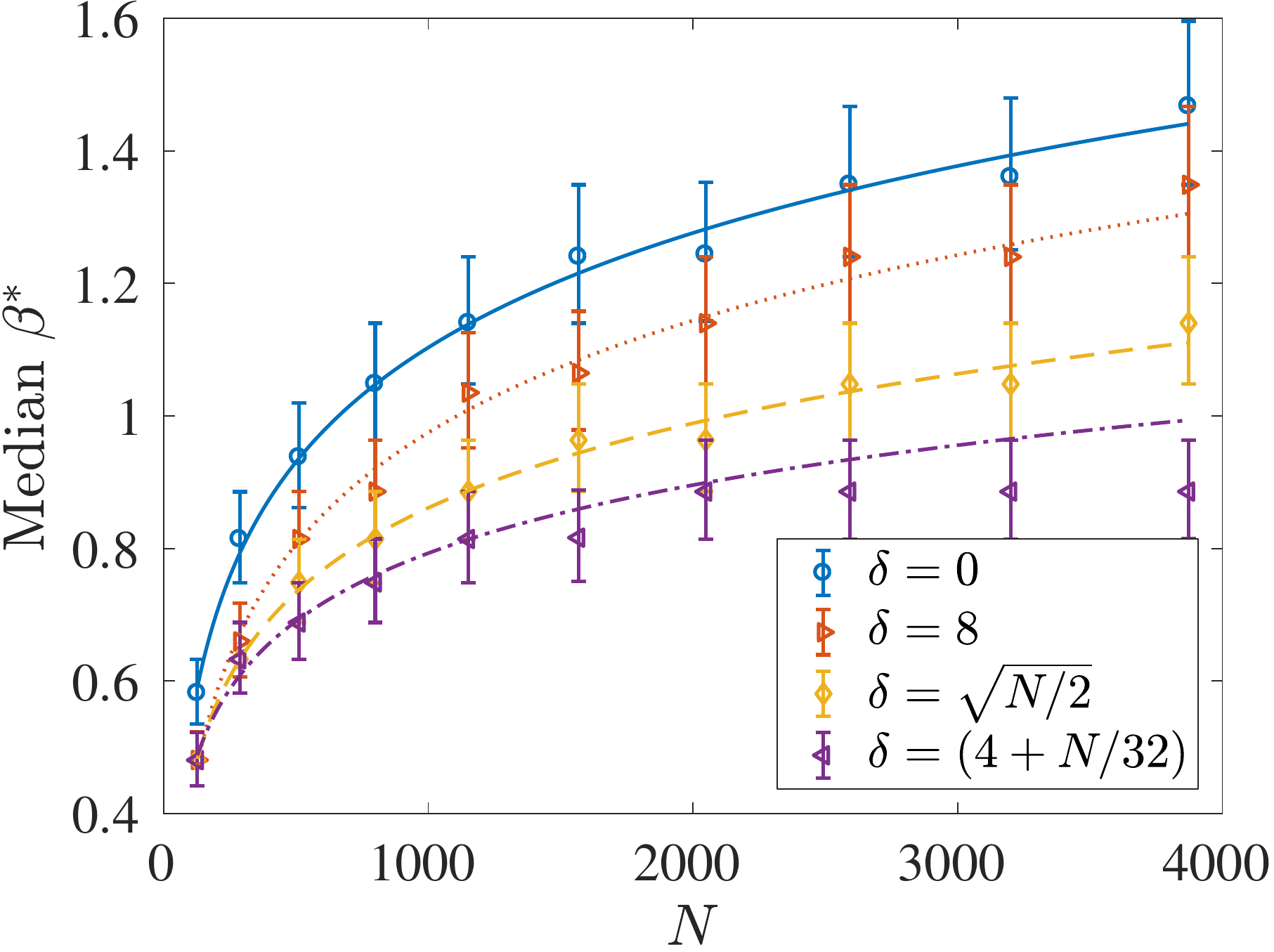} 
   \caption{{\bf The behavior of the median value (over 100 instances) of the minimum inverse-temperature required such that for $q = 10^{-3}$ the target energy is the ground state.}  Error bars correspond to the spacing between the $\beta$ values of the PT simulations.  Lines correspond to the fits $\beta = a + b \ln N$ with $ b= 0.2495 \pm 0.0531, 0.2440 \pm 0.0461, 0.1842 \pm 0.0423, 0.1483 \pm 0.1274$ for $\delta = 0, 8, \sqrt{N/2}, (4+N/32)$ respectively, with the uncertainty representing the 95\% confidence interval for the fit parameters.}
   \label{fig:plantedTarget}
\end{figure}

In Fig.~\ref{fig:plantedTarget}, we supplement the results presented in the main text with the scaling of $\beta$ when the target energy need not be the ground state, specifically $E_{\mathrm{T}} = E_0 + \delta$.  We consider three  cases: (i) a constant about the ground state, $E_{\mathrm{T}} = E_0 + 8$, (ii) a square-root scaling above the ground state, $E_{\mathrm{T}} = E_0 + \sqrt{N/2}$, and a linear scaling above the ground state $E_{\mathrm{T}} = E_0 + (4+N/32)$.  The specific values were picked so that the three cases would have the same target energy at the smallest size of $N=128$.  If we fit all curves with a logarithmic dependence on $N$, we observe a similar scaling for the cases of $\delta = \mathrm{constant}$, and the case of $\delta \propto \sqrt{N}$ still exhibits a logarithmic scaling but with a milder coefficient.  For the case of $\delta \propto N$, the required $\beta$ approaches a constant for sufficiently large problem sizes.

\section{Results for the 3-reg 3XORSAT and random $\pm 1$ Chimera Instances} \label{sec:Additional}
\begin{figure}[htbp] 
   \centering
   \subfigure[]{\includegraphics[width=0.9\columnwidth]{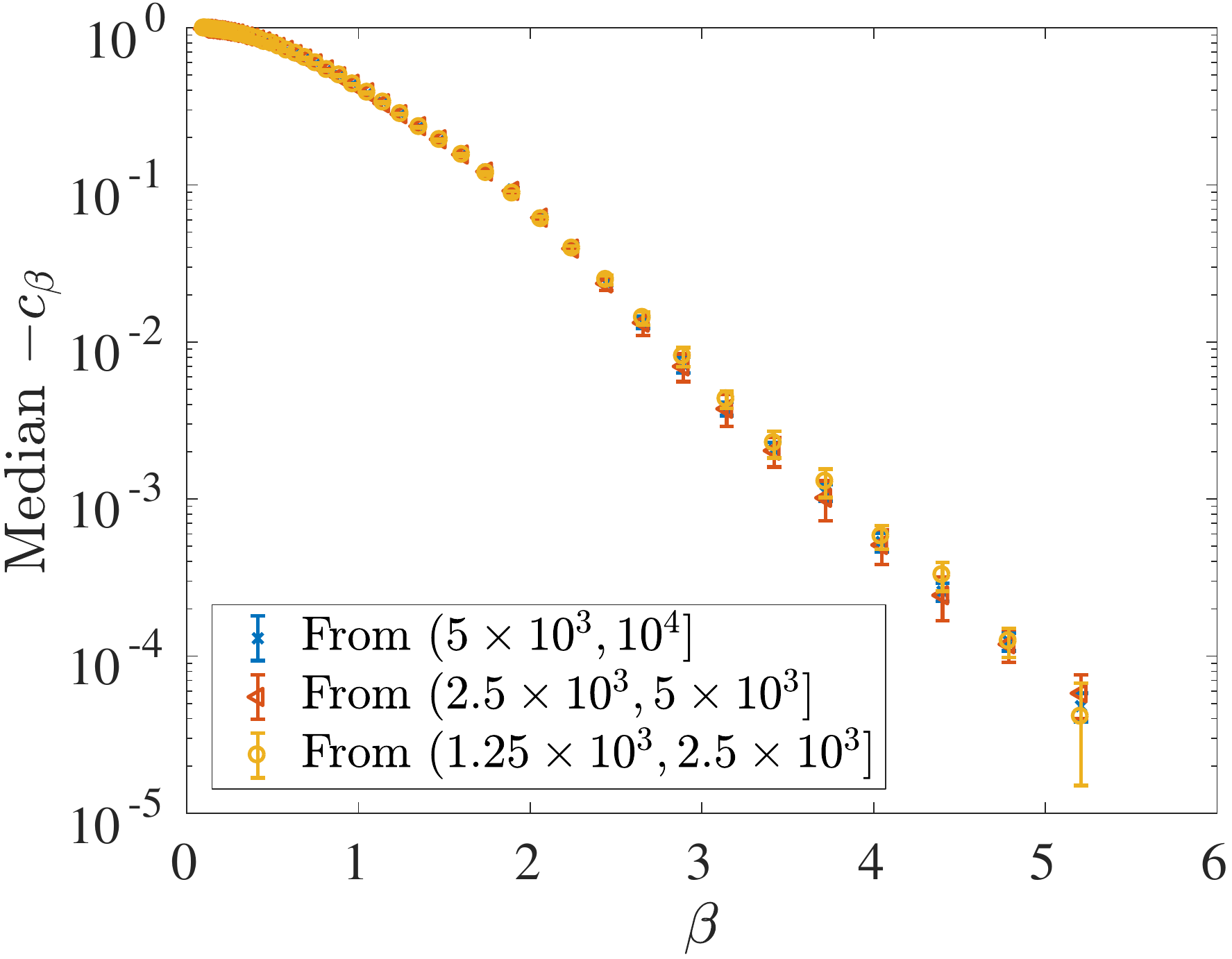}}
   \subfigure[]{\includegraphics[width=0.9\columnwidth]{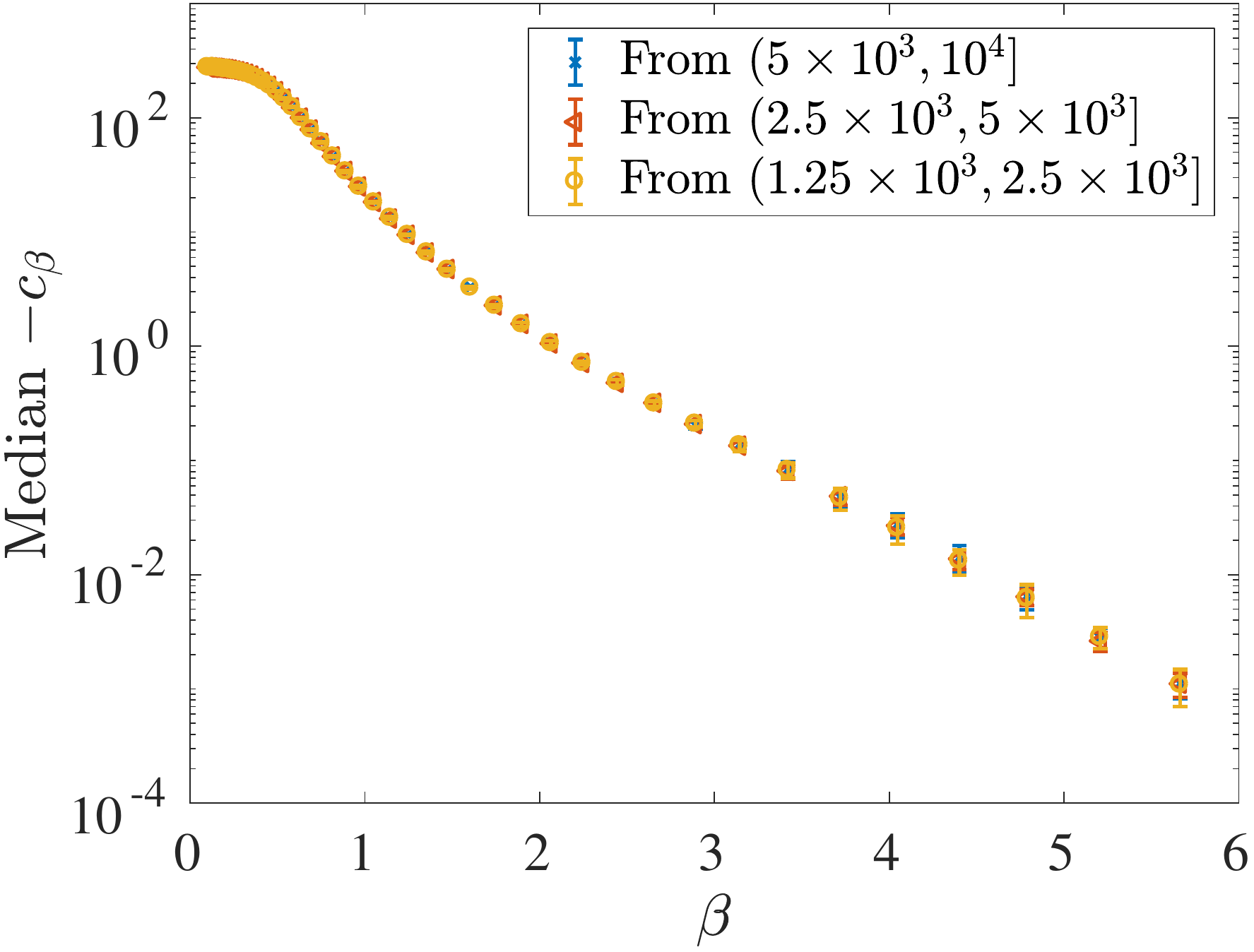}}
   \caption{{\bf The behavior of the median specific heat for the (a) 3-regular 3-XORSAT instances ($N=100$) and (b) the bimodal instances ($L=12$) using different blocks of the samples}.  Of the total of $10^4$ samples, different partitions (as indicated by the legend) are used to calculate the specific heat.}
   \label{fig:XORSATTest}
\end{figure}

\begin{figure}[thbp] 
   \centering
   \includegraphics[width=0.95\columnwidth]{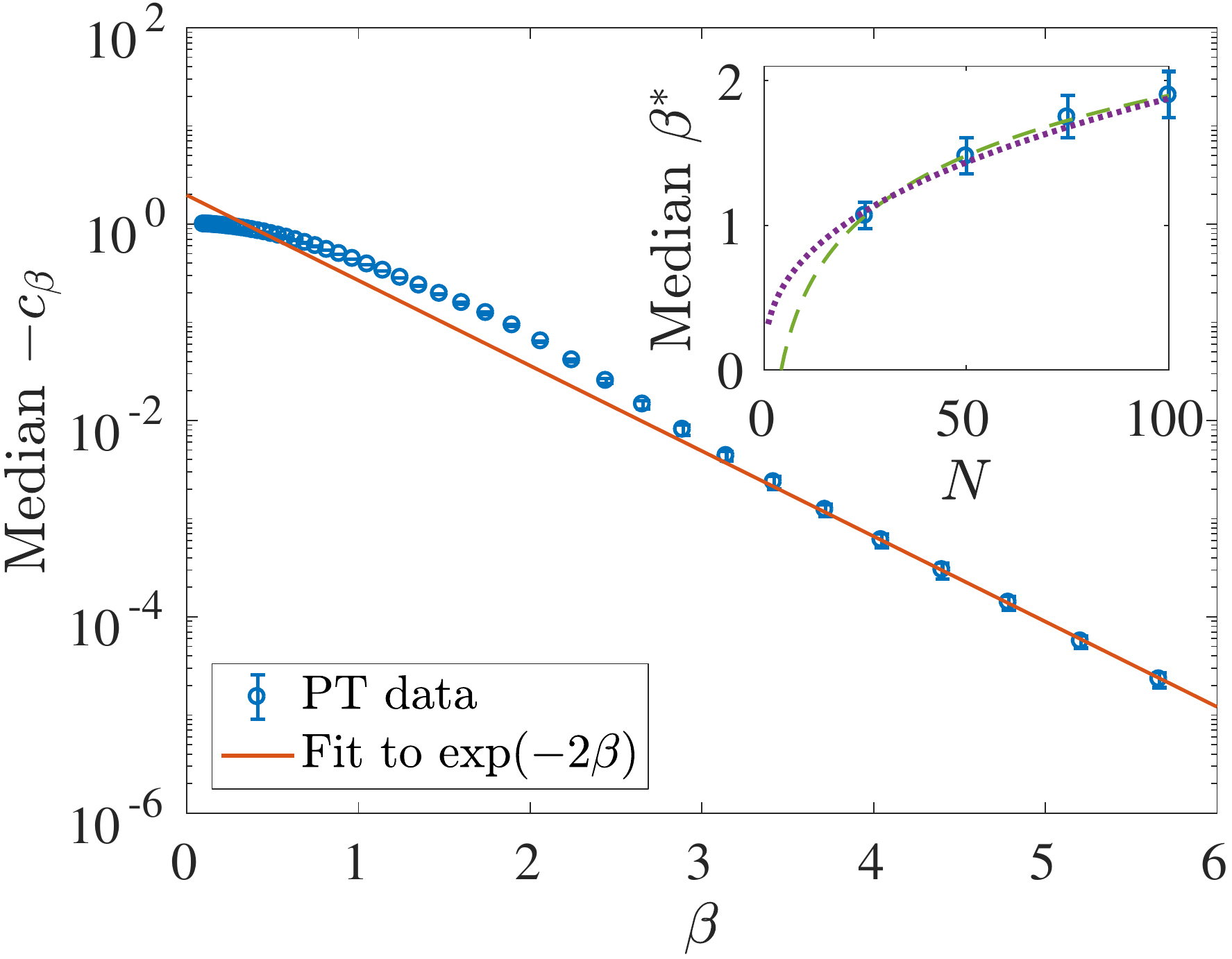}
   \caption{{\bf Behavior of the median specific heat (over 100 instances) for the 3-regular 3-XORSAT instances with inverse-temperature $\beta$ for $N =100$}.  The behavior transitions from a polynomial scaling with $\beta$ to an exponential scaling with $\beta$. {\bf Inset:} Typical minimum inverse-temperature required for instances of size $N$ such that probability of the ground state is at least $q = 10^{-1}$.  Also shown are fits to $\log N$ anto d a power-law $c N^{\alpha}$ with $\alpha=0.39\pm0.18$, which is almost indistinguishable from the logarithmic fit for large size.}
   \label{fig:XORSATBetaTarget}
\end{figure}

\begin{figure}[htbp] 
   \centering
   \includegraphics[width=0.95\columnwidth]{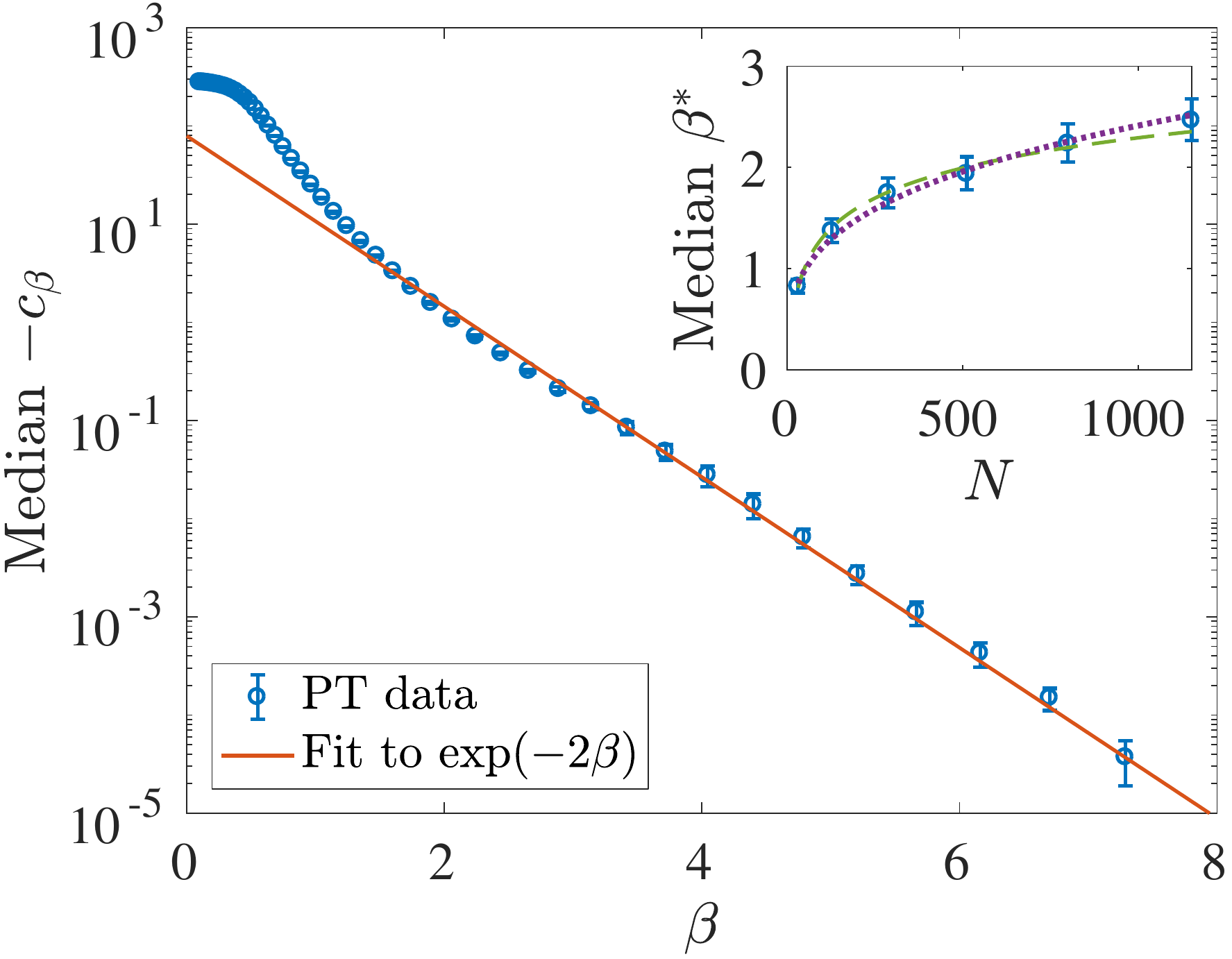}
   \caption{{\bf Behavior of the median specific heat (over 100 instances) for the random $\pm 1$ instances with inverse-temperature $\beta$ for $N =1152$}.  The behavior transitions from a polynomial scaling with $\beta$ to an exponential scaling with $\beta$. {\bf Inset:} Typical minimum inverse-temperature required for instances of size $N$ such that probability of the target energy $E_{\mathrm{T}} = E_{0} + \delta(N)$ is at least $q = 10^{-1}$.  
 Also shown are fits to $\log N$ for all three cases and to a power-law $c N^{\alpha}$ with $\alpha=0.30\pm0.09$ , which is almost indistinguishable from the logarithmic fit for large sizes.}
   \label{fig:BiModalBetaTarget}
\end{figure}

Here we provide the equivalent plots to Fig.~(3) of the main text but for the 3-regular 3-XORSAT (Fig.~\ref{fig:XORSATBetaTarget})  and random $\pm 1$ instances (Fig.~ \ref{fig:BiModalBetaTarget}).  The random $\pm 1$ instances were warmed-up with up to $24\times 10^6$ PT swaps depending on their size, while the XORSAT instances were warmed-up for with up to $200 \times 10^6$ swaps depending on their size.  For both, as in the planted-solution case, $10^4$ samples were taken with one sample after every 50 PT swaps.  We perform the same thermalization test as for the planted-solution instances, and we observe no significant difference for the different blocks of samples (see Fig.~\ref{fig:XORSATTest}).

We note that for both of these classes of instances, the $\beta$ values required fall in the regime where the scaling of the specific heat with $\beta$ is not yet exponential.  The scaling behavior of $\beta^\ast$ is consistent with both a $\log N$ and a $N^{1/\alpha}$ behavior.
%
%

\section{Scaling laws for temperatures: analytical examples} \label{sec:Analytics}
Let us consider the simple case of non-interacting spins in a global magnetic field.  This case is particularly relevant if the initial state of the quantum annealer is prepared as the thermal state of the standard driver Hamiltonian $-\frac{1}{2} \sum_{i=1}^N \sigma_i^x$ with no overall energy scaling.  The partition function is given by:
\beq
Z = \sum_{k=0}^N {N \choose k } e^{- \beta \left( k - N/2 \right)} = \left[2 \cosh(\beta/2) \right]^N 
\eeq
Note that each energy spectrum has a degeneracy that grows polynomially with $N$.  
The mean energy is given by:
\beq
\mu/N = - \frac{1}{2} \tanh(\beta/2)
\eeq
and the standard deviation is:
\beq
\sigma/ \sqrt{N} = \frac{1}{2} \mathrm{sech}(\beta/2)
\eeq
The ground state probability on a thermal state is then given by $p_\GS = \frac{e^{\beta N /2}}{Z}$, which we can then invert to write the inverse-temperature as:
\beq
\beta = - \ln \left( 1 - p_\GS^{-1/n} \right)
\eeq
If we pick $p_\GS$ to be some small but fixed (independent of system size) number and take the large $N$ limit, we find that
\beq
\beta = \ln(N) - \ln(- \ln p_\GS) + \frac{1}{2N} \ln p_\GS + \dots
\eeq
Therefore, we find that for this simple problem, in order to maintain a constant ground state probability while the system size grows, we must scale the inverse-temperature logarithmically with system size.

A Grover search problem~\cite{Grover:97a,Roland:2002ul} on the other hand yields the worst case scaling. In this case, we take a single state to have energy $-N$, while the remaining states have energy $-N+1$.  The partition function is given by:
\beq
Z = e^{\beta N} \left[1 + (2^{N} - 1 ) e^{-\beta} \right]
\eeq
with mean energy: 
\beq
\mu =  -N+1 - \frac{e^{\beta}}{2^N - 1 + e^{\beta} }
\eeq
and standard deviation
\beq
\sigma = e^{\beta/2} \frac{\sqrt{2^N - 1}}{2^N -1 + e^{\beta}}
\eeq
Unlike our other local example the $\sigma$ does not scale as $\sqrt{N}$.  The ground state probability is given by $p_\GS = 1/Z$.  Inverting this for $\beta$, we find:
\beq
\beta = \ln(2^N - 1) - \ln \left( p_\GS^{-1} -1 \right)
\eeq
Again, for a fixed and small $p_\GS$, expanding for large $N$, we get:
\beq
\beta = N \ln 2 - \ln \left( p_\GS^{-1} -1 \right) - 2^{-N} + \dots
\eeq
Therefore, in this case, $\beta$ must grow linearly with $N$ in order to maintain a constant $p_\GS$.
Note of course that the Grover Hamiltonian is highly non-local as it contains $N$-body terms.
 \section{Final ground state weight in Gibbs distributions}
 We consider here the weight of the final ground state on the Gibbs distributions along a quantum annealing protocol.  Let us consider a system of $N$ decoupled qubits evolving under:
 \beq
 H(s) = -\frac{1}{2} (1-s) \sum_{i=1}^N \sigma_i^x - \frac{1}{2} s \sum_{i=1}^N \sigma_i^z
 \eeq
 The probability of the final ground state, the state $\ket{0}^{\otimes N}$, in the instantaneous Gibbs distribution is given by:
 \beq
 p(s) = \left( \frac{ | \braket{0}{\lambda_0(s)}|^2 e^{\beta \lambda(s)/2} + | \braket{0}{\lambda_1(s)}|^2 e^{-\beta \lambda(s)/2} }{ 2 \cosh(\beta \lambda(s)/2)} \right)^N
 \eeq
 where $\ket{\lambda_0(s)}$ and $\ket{\lambda_1(s)}$ are the instantaneous ground state and first excited state for the single qubit system with eigenvalues $-\lambda(s)$ and $\lambda(s)$ respectively.  Let us define $\Lambda(s) \equiv | \braket{0}{\lambda_0(s)}|^2 = 1- | \braket{0}{\lambda_1(s)}|^2$, so we can rewrite our expression as:
 \beq
p(s) = \Lambda(s) \tanh( \beta \lambda(s) /2) + \frac{1}{ 1 + e^{\beta \lambda(s)}}
 \eeq
 We therefore have:
 \beq
 \frac{d}{ds} p(s) = \frac{\Lambda'(s) \left( e^{2 \beta \lambda(s)} - 1 \right) + \left( 2 \Lambda(s) -1 \right) \beta \lambda'(s) e^{\beta \lambda(s)}}{\left(1+e^{\beta \lambda(s)} \right)^2}
 \eeq
 Note that $\Lambda'(s) > 0 , 2 \Lambda -1 > 0\forall s$, and $  \lambda'(s) < 0$  for $s < 0.5$.  We can therefore ask, is it possible for  $  \frac{d}{ds} p(s) = 0$ for $s < 0.5$?  Because of the exponential factors, for large $\beta$ the first term will dominate the second term, and this will not occur.  Let us therefore consider the small $\beta$ case.  If we expand the exponentials, we find 
 \beq
  \frac{d}{ds} p(s) = \frac{\beta}{4} \left( 2 \Lambda'(s) \lambda(s) + (2 \Lambda(s) - 1) \lambda'(s) \right) + O(\beta^2)
 \eeq
 However, an explicit evaluation of the expression in parenthesis gives 1, i.e.:
 \beq
 2 \Lambda'(s) \lambda(s) + (2 \Lambda(s) - 1) \lambda'(s) = 1
 \eeq
 Therefore, even in the high temperature limit, $ \frac{d}{ds} p(s) $ remains positive.   Numerically, we can confirm that  $ \frac{d}{ds} p(s) $ remains positive.  Therefore, we can conclude that $p(s)$ is monotonically increasing and achieves its maximum value at $s = 1$.
 
 For the Grover problem, we take \cite{Roland:2002ul}
 \beq
 H(s) = (1-s) \left( 1 - \ketbra{\phi}{\phi} \right) + s \left( 1 - \ketbra{m}{m} \right)
 \eeq
 where $\ket{\phi}$ is the uniform superposition state and $\ket{m}$ denotes the `marked' state which is the ground state at $s = 1$.  The spectrum is such that only the instantaneous ground state and first excited state have non-zero weight on the marked state for $s<1$.  These two states can be written as:
 \bes
 \begin{align}
 \ket{\lambda_0(s)} & =  \cos \frac{\theta(s)}{2} \ket{m} + \sin \frac{\theta(s)}{2} \ket{m^{\perp}} \ , \\
\ket{\lambda_1(s)}  &=  -\sin \frac{\theta(s)}{2} \ket{m} + \cos \frac{\theta(s)}{2} \ket{m^{\perp}} \ .
 \end{align}
 \ees
 with eigenvalues $\frac{1}{2} (1 - \Delta(s))$ and $\frac{1}{2} (1 + \Delta(s))$ respectively and
 \bes
\begin{align}
\label{eq:gap-Grover}
\Delta(s) & =  \sqrt{ (1-2 s)^2 + \frac{4}{2^N} s (1-s) } \ , \\
\cos \theta(s) & =  \frac{1}{\Delta(s)} \left[ 1 - 2 (1-s) \left( 1 - \frac{1}{2^N} \right) \right] \ , \label{eqt:cosGrover}\\
\sin \theta(s) & =  \frac{2}{\Delta(s)} \left(1 - s \right)  \frac{1}{\sqrt{2^N}}  \sqrt{1 - \frac{1}{2^N}} \ .
\end{align}
\ees
The probability of the final ground state in the instantaneous Gibbs distribution is given by:
 \beq
 p(s) = \frac{\cosh(\beta \Delta(s)/2) + \cos \theta(s) \sinh(\beta \Delta(s)/2)}{2 \cosh(\beta \Delta(s)/2) + (2^N-2)e^{-\beta/2}}
 \eeq
For small $\beta$, one has:
\beq
\frac{d}{ds} p(s) = \frac{\beta}{2^{N+1}} \frac{d}{ds} \left( \Delta(s) \cos \theta(s) \right) + O(\beta^2)
\eeq
and it is clear from Eq.~\eqref{eqt:cosGrover} that this expression is positive for all $s$.   Numerically, we can confirm that  $ \frac{d}{ds} p(s) $ remains positive for $s\in[0,1]$.  Therefore, we can conclude that $p(s)$ is monotonically increasing and achieves its maximum value at $s = 1$.

\end{document}